\input iopppt
\pptstyle
\title{Some Restrictions on the Symmetry Groups of Axially
Symmetric Spacetimes}[Symmetries in Axially Symmetric Spacetimes]

\author{Alan Barnes}

\address{Computer Science Group, School of Engineering and Applied Science,
Aston University, Aston Triangle, Birmingham B4 7ET, UK
\footnote\dag{E-Mail: \ \ {\tt barnesa{\bf @}aston.ac.uk}
\ \ \ Fax:\ \  {\rm +44 121 333 6215}}}

\jl{6}
\submitted

\date

\beginabstract
Lie transformation groups containing a one-dimensional
subgroup acting cyclically on a manifold are considered.  The
structure of the group is found to be considerably
restricted by the existence of a one-dimensional subgroup whose orbits
are circles. The results proved do not depend on the dimension of the
manifold nor on the existence of a metric, but merely on the fact that
the Lie group acts globally on the manifold.
Firstly some results for the general case of an $m+1$-dimensional Lie
group are derived: those commutators of the associated Lie algebra
involving the generator of 
the cyclic subgroup, $X_0$ say, are severely restricted and, in a
suitably chosen basis, take a simple form. The Jacobi identities
involving $X_0$ are then applied to show there are further
restrictions on the structure of the Lie algebra. All Lie algebras of
dimensions 2 and 3 compatible with cyclic symmetry are obtained. In
the two-dimensional case the group must be Abelian. For 
the three-dimensional case, the Bianchi type of the Lie algebra must
be I, II, III, VII$_0$, VIII or IX and furthermore in all cases
the vector $X_0$ forms part of a basis in which the algebra takes its
canonical form.  Finally four-dimensional Lie algebras compatible
with cyclic symmetry are considered and the results are related to the
Petrov-Kruchkovich classification of all four-dimensional Lie
algebras.
\endabstract
\vfill\eject

\section{Introduction}
Carter (1970) defined a spacetime $\cal M$ to admit a  
cyclical symmetry if and only if the metric is invariant under the
effective smooth action $SO(2) \times \cal M \rightarrow \cal M$ of the
one-parameter cyclic group $SO(2)$.
A cyclically symmetric spacetime
in which the set of fixed points of this isometry is not empty is said
to be axially symmetric and the set of fixed points itself is referred
to as the axis (of symmetry).
Cyclic symmetry is thus a slight generalisation of
the concept of axial symmetry.  Although the assumption of the
existence of an axis is reasonable in many circumstances,
there are physically realistic situations where
an axis may not exist:  the 'axis' may be singular and so not part of the
manifold, or the topology of the manifold may be such that no axis
exists. An example of the latter situation is a torus embedded in a
3-dimensional flat space; the torus is cyclically symmetric but the
Killing vector field generating the rotation does not vanish anywhere
on the surface. 

Mars and Senovilla (1993) proved a
number of useful results on the structure of the axis if it exists.
Carot, Senovilla and Vera (1999) showed in an axial symmetric
spacetime that if there is a second Killing vector which, with the Killing
vector generating the axial symmetry, generates a two-dimensional isometry
group then the two Killing vectors commute and the isometry group is
Abelian.  A similar result for stationary axisymmetric spacetimes was
proved by Carter (1970). The proofs of both results depend heavily on
the existence of an axis.  It is interesting therefore to consider whether a
result of the same ilk as those of Carter and Carot et al.\ are valid for
cyclically symmetric spacetimes.   Bi\v c\'ak \& Schmidt (1984) showed
that if in cyclically symmetric spacetime  there is a second Killing
vector which, with the Killing vector generating the cyclic symmetry,
generates a two-dimensional isometry group then the two Killing
vectors commute.
As pointed out by Barnes (2000) this result does not depend on the
dimension of the manifold, nor on the fact that the transformation
group is an isometry group nor on the existence of a non-empty axis
Thus the following propostion is valid:
any two-dimensional Lie transformation group acting on a manifold
which contains a one-dimensional subgroup whose orbits are circles,
must be Abelian. 
The method used to prove this result was extended by Barnes (2000) to
apply to three-dimensional Lie transformation groups and it was shown
that the existence of a one-dimensional subgroup with closed orbits
restricts the Bianchi type of the associated Lie algebra to be  I, II, III,
VII$_0$, VIII or IX.   

In this paper we will consider Lie transformation groups of arbitrary
(finite) dimension containing a one-dimensional subgroup acting
cyclically on the manifold (that is with closed circular orbits) and
show that the structure of the group is restricted considerably by this
assumption.  The term ``cyclically symmetric manifold'' will be used
to refer to this situation which is thus being used in a less
restricted sense than that of Carter (1970) which referred only to
isometry groups.
Therefore the results below may be applied not just to isometries, but to
more general symmetries: homothetic and conformal motions, affine,
projective and curvature collineations etc.. Furthermore the results
do not depend the existence of an axis nor on the dimension of the
manifold nor indeed on the existence of a metric. 

\section{Cyclically symmetric manifolds admitting a $G_{m+1}$}
Suppose $\cal M$ is an $n$ dimensional manifold which admits an $m+1$
dimensional Lie transformation group $G_{m+1}$ acting on $\cal M$
which contains a one-dimensional subgroup isomorphic to $SO(2)$.  
Suppose $X_0$ is the vector field generating this subgroup.
We restrict attention to the open submanifold $\cal N$ of $\cal M$ on
which $X_0$ is non-zero. 
The orbit each point of $\cal N$  under the cyclic symmetry is a
circle and these circles are the integral curves of the vector
field $X_0$. 
Let $\phi$ be a circular coordinate running from $0$ to $2\pi$ which
labels the elements of $SO(2)$ in the usual way.  Then we can
introduce a system of coordinates 
$x^i$ with $i = 1 \ldots n$ and $x^1 = \phi$  adapted to $X_0$ such that
$X_0 = \partial_\phi$.  These coordinates are determined only up to
transformations of the form
$$\tilde \phi = \phi + f(x^\nu) \qquad {\tilde x}^\mu = g^\mu(x^\nu)$$
where $f$ and $g^\mu$ are smooth functions and where Greek
indices take values in the range $2 \ldots n$.

Let $X_a$ be vector fields on $\cal M$ which, together with $X_0$,
form a basis of the Lie algebra of $G_{m+1}$.  Here and below indices
$a$, $b$ and $c$ take values in the range $1\ldots m$.
The commutators involving $X_0$ may be written in the form
$$[X_0\ X_a] = A^b_a X_b + B_a X_0 \en $$
where $A^b_a$ and $B_a$ are constants and the Einstein summation
convention has been used over the range $1 \ldots m$.
If we introduce new basis vectors $\tilde X_a$ given by
$\tilde X_a = P^b_a X_b$, where $P^b_a$ are constants, the structure
constants transform as follows 
$$\tilde {\bf A} = {\bf P} {\bf A} {\bf P}^{-1}
\qquad \qquad \tilde {\bf B} = {\bf P}{\bf B}\en $$
where for conciseness standard matrix notation has been used.
Using these transformations the matrix ${\bf A}$ may be reduced to Jordan
normal form.  In what follows we will work in a basis in which the structure
constants $A^b_a$ are in Jordan normal form but, for typographic
simplicity, tildes will be omitted.  

Let $X$ be a Jordan basis vector which  is a real eigenvector of
$\bf A$ with corresponding  real eigenvalue $\lambda$, the commutation
relation (1) for $X$ is thus
$$[X_0\ X] = \lambda X + B X_0 \en $$
where $B$ is a constant.  In terms of a coordinate system  adapted to
$X_0$ in which  $X_0 =  \partial_\phi$, the commutation relation becomes
$$ {\partial X^i \over \partial \phi} = \lambda X^i +B\delta^i_0$$
where $i$ ranges over $1\ldots n$.  If the solutions of these
equations are to be periodic in $\phi$,  then the eigenvalue $\lambda$
must zero and moreover $B = 0$ in (3).

Suppose now that $Z = X + iY$ (where $X$ and $Y$ are real vectors) is an
eigenvector of ${\bf A}$ with corresponding eigenvalue $\lambda + i
\mu$, the commutation relation (1) for $Z$ reduces to
$$[X_0\ X] =\lambda X - \mu Y + B X_0 \qquad 
[X_0\ Y] =\lambda Y + \mu X + C X_0 \en $$
where $B$ and $C$ are real constants.  In terms of an adapted
coordinate system in which  $X_0 =  \partial_\phi$, these equations
reduce to 
$${\partial X^i \over \partial \phi} = \lambda X^i - \mu Y^i
+B\delta^i_0
\qquad {\partial Y^i \over \partial \phi} = \lambda Y^i + \mu X^i
+C\delta^i_0$$
where $i$ takes values in the range $1 \ldots n$.
If the solutions of these equations are to be periodic in $\phi$ with
period $2\pi$, then $\lambda = 0$ and $\mu = \pm N$ where $N$ is a
positive integer. Furthermore by redefining $X$ and $Y$ according to 
$$\tilde X = X + C\mu^{-1}X_0 \qquad \tilde Y = Y - B\mu^{-1}X_0$$
we may set $B=C=0$ in the commutation relations (4).

Thus we have shown that the only eigenvalues of ${\bf A}$ are 
zero or of the form $\pm i N$ where $N$ is an integer. Furthermore all of
the Jordan blocks must be simple or equivalently the minimal polynomial
of ${\bf A}$ must have no repeated factors. To see this suppose
first that there is a non-simple elementary divisor with eigenvalue
zero. Thus there is a vector $X$ such that ${\bf A}^2 X = 0$ but 
${\bf A} X  \equiv Y \ne 0$. Thus we have 
$$[X_0\ X] = Y + BX_0 \qquad [X_0\ Y] = 0$$
where $B$ is a constant. In a coordinate system in which 
$X_0 =  \partial_\phi$, these reduce to 
$${\partial X^i \over \partial \phi} = Y^i +B\delta^i_0
\qquad {\partial Y^i \over \partial \phi} = 0$$
The solution $X$ of these differential equations is clearly linear in
$\phi$ as $Y \ne 0$. However the components of the vectors $X$ and $Y$
must be periodic in $\phi$, which is a contradiction. A
similar argument shows that there can be no non-simple elementary
divisors of ${\bf A}$ corresponding to the purely imaginary
eigenvalues.

Thus without loss of generality, we may suppose that ${\bf A}$ has $p$ 
eigenvalues for some $p$ with $0 \le  2p \le m$ of the form $i n_j$
where each $n_j$ is a positive integer and $1 \le j \le p$ with
corresponding complex eigenvectors  
$Z_j \equiv  X_{2j} + i X_{2j-1}$ plus $m-2p$ zero eigenvalues with
corresponding real eigenvectors $X_k$  where $2p+1 \le k \le m$.  Choosing
these $m$  vectors $X_a$ as the basis vectors, the commutators
become 
$$\eqalignno{
[X_0\ X_{2j-1}] = n_j X_{2j}\qquad  & {\rm for\ }1 \le j \le p 
\enpt\cr
[X_0\ X_{2j}] = -n_j X_{2j-1}\qquad  & {\rm for\ }1 \le j \le p  
\enpt \cr
[X_0\ X_k] = 0 \qquad & {\rm for\ }2p+1 \le k \le  m  \endpt
}$$
If the $n_j$'s are distinct, the remaining basis freedom preserving
the commutation relations (5) is a rotation and dilatation in each of
the $p$ two-planes generated by the pairs of vectors $X_{2j-1}$ and
$X_{2j}$ for $1 \le j \le p$ plus a general (non-singular) transformation
of the vectors $X_k$ for $2p+1 \le k \le m$:
$$\eqalignno{
\tilde X_{2j-1} & = C_j(\cos \theta_j X_{2j-1} + \sin \theta_j X_{2j})
 \enpt \cr
\tilde X_{2j} & = C_j(-\sin \theta_j X_{2j-1} + \cos \theta_j X_{2j}) 
 \enpt \cr
\tilde X_k & = A^l_k X_l  + B_k X_0  \endpt
}$$
where $A^l_k$ and $B_k$ are constants for $k,l = 2p+1 \ldots m$, as
are $C_j$ and $\theta_j$ for $j = 1 \ldots p$.
If the $n_j$'s are not all distinct there is some additional 
freedom in the choice of the associated pairs of basis vectors.

We now consider the restrictions imposed on the commutators by those
Jacobi identities which involve the generator $X_0$ of the cyclic
subgroup
$$[X_0\ [X_i\ X_j]] + [X_i\ [X_j\ X_0]] +[X_j\ [X_0\ X_i]] = 0 \en $$
where $i,j = 1 \ldots m$.  If $p=0$, that is if all the vectors
commute with $X_0$, these Jacobi identities yield no information. For
$p \neq 0$ there are four cases to consider as follows.

(i) If $X_i$ and $X_j$ both commute with $X_0$, we have $[X_0\ [X_i\ X_j]] = 0$
and hence the structure constants satisfy
$$C^k_{ij} = 0\qquad {\rm for\ } 1 \le k \le 2p \quad {\rm and}  \quad
2p+1 \le i,j \le m \en $$

(ii) If $X_{2k-1}$ and $X_{2k}$ are the basis vectors associated with the
imaginary eigenvalue $in_k$,
then we have $[X_0\ [X_{2k-1}\ X_{2k}]] = 0$ and thus
$$C^i_{2k-1\,2k} = 0\qquad {\rm for}\quad 1 \le i \le 2p \quad {\rm and}
\quad 1 \le k \le p \en $$

(iii) If $X_j$ commutes with $X_0$ but $X_i$ does not, then we have 
$$\eqalignno{
[X_0\ [X_{2k-1}\ X_j]] = n_k [X_{2k}\ X_j] \qquad 
 &{\rm for\ } i =2k-1 \cr
[X_0\ [X_{2k}\ X_j]] = -n_k [X_{2k-1}\ X_j] \qquad 
&{\rm for\ } i = 2k}
$$
where $1 \le k \le p$ and  $2p+1 \le j \le m$.  Thus immediately we can deduce
$$C^0_{ij} = C^l_{ij} = 0 \qquad 
{\rm for\ } 1\le i \le 2p,\ 2p+1 \le j,l \le m \en $$
Also we have
$$\eqalignno{
n_l C^{2l-1}_{2k-1\,j} = n_kC^{2l}_{2k\,j} \qquad
& -n_lC^{2l}_{2k-1\,j} = n_kC^{2l-1}_{2k\,j}  \enpt\cr
n_l C^{2l-1}_{2k\,j} = -n_kC^{2l}_{2k-1\,j} \qquad
& -n_lC^{2l}_{2k\,j} = -n_kC^{2l-1}_{2k-1\,j}  \endpt
}$$
where $1 \le k,l \le p$ and  $2p+1 \le j \le m$.
Thus 
$$C^{2k-1}_{2k-1\,j} =C^{2k}_{2k\,j} \qquad 
C^{2k}_{2k-1\,j} =C^{2k-1}_{2k\,j} \en$$
for $1 \le k \le 2p$ and $2p+1 \le j \le m$.
For $k \neq l$, eliminating $C^{2l-1}_{2k-1\,j}$ and
$C^{2l-1}_{2k\,j}$ from (11) we have
$$(n_l^2 - n_k^2)C^{2l}_{2k-1\,j} =0 \qquad (n_l^2 - n_k^2)C^{2l}_{2k\,j} =0$$
Since all $n_i$'s are positive,  if $n_l \neq n_k$  we have  
$$C^{2l-1}_{2k-1\,j} = C^{2l-1}_{2k\,j} = C^{2l}_{2k-1\,j} = 
 C^{2l}_{2k\,j} =0 \qquad {\rm for} \quad l \ne k \en $$ 
where $1 \le k,l \le p$ and  $2p+1 \le j \le m$.

(iv)  If the basis vectors are associated with different imaginary
eigenvalues $n_i$ and $n_j$ then 
$$\eqalignno{
[X_0\ [X_{2i-1}\ X_{2j-1}]] &= n_j[X_{2i-1}\ X_{2j}] +n_i[X_{2i}\ X_{2j-1}]
\enpt \cr
[X_0\ [X_{2i-1}\ X_{2j}]] &= -n_j[X_{2i-1}\ X_{2j-1}] +n_i[X_{2i}\ X_{2j}]
\enpt \cr
[X_0\ [X_{2i}\ X_{2j-1}]] &= n_j[X_{2i}\ X_{2j}] -n_i[X_{2i-1}\ X_{2j-1}]
\enpt \cr
[X_0\ [X_{2i}\ X_{2j}]] &= -n_j[X_{2i}\ X_{2j-1}] -n_i[X_{2i-1}\ X_{2j}] 
\endpt }$$
It follows immediately that 
$$\eqalignno{
n_jC^k_{2i-1\,2j} + n_iC^k_{2i\,2j-1} =0 \qquad
&-n_jC^k_{2i-1\,2j-1} + n_iC^k_{2i\,2j} =0 \enpt \cr
n_jC^k_{2i\,2j} - n_iC^k_{2i-1\,2j-1} =0 \qquad
&-n_jC^k_{2i\,2j-1} - n_iC^k_{2i-1\,2j} =0 \endpt
}$$
If the integers $n_i$ are distinct we may deduce
$$C^k_{2i-1\,2j-1}=C^k_{2i-1\,2j}=C^k_{2i\,2j-1} = C^k_{2i\,2j}=0 \en $$
for $1 \le i,j \le p$ and $k=0$ or $2p+1 \le k \le m$ with  $i \neq j$.
Also we obtain two sets each of four linear equations for four structure
constants:

$$\left(\matrix{n_h &0 &-n_i &-n_j \cr
	      0   &n_h &n_j  &n_i  \cr
	      -n_i  &n_j &n_h  &0  \cr
              -n_j  &n_i &0    &n_h}\right)
\left(\matrix{C^{2h-1}_{2i-1\,2j-1}\hfill\cr
              C^{2h-1}_{2i\,2j} \hfill\cr
              C^{2h}_{2i-1\,2j} \hfill\cr
              C^{2h}_{2i\,2j-1} \hfill}\right) = {\bf 0}\enpt$$
$$\left(\matrix{n_h &0 &n_i &-n_j \cr
                0  &n_h &n_j  &-n_i  \cr
                n_i  &n_j &n_h  &0  \cr
               -n_j  &-n_i &0   &n_h}\right)
\left(\matrix{C^{2h-1}_{2i-1\,2j}\hfill\cr
              C^{2h-1}_{2i\,2j-1}\hfill\cr
              C^{2h}_{2i-1\,2j-1}\hfill\cr
              C^{2h}_{2i\,2j}\hfill}\right) = {\bf 0}\endpt$$
where $1 \le h,i,j \le p$ with $i \neq j$ . 
If $n_h \neq \pm n_i \pm n_j$,
that is if no integer $n_k$ is a sum of two other such integers, the
matrices in (17a,b) are non-singular and it
follows that all of these eight structure constants are zero. Thus with the
aid of equation (9) we may deduce
$$C^k_{ij}= 0 \qquad {\rm for} \quad 1 \le i,j,k \le 2p \en $$

Thus for $p>0$, we have shown that the Jacobi identities imply that a number of
blocks of the structure constants vanish as given by equations (8), (9) and
(10) and in addition equation (12) is always satisfied.  Also
for the generic case where the $p$ integers $n_i$ are distinct and no
integer $n_i$ is the sum of two other such integers even more of the
structure constants vanish as given by Eqs. (13), (16) and (18). 

Even if the case where $n_i=n_j$  for 
some $i \neq j$, there are still restrictions on the structure
constants imposed by the Jacobi identities; the solutions
of equations (11) and (15) are respectively
$$C^{2l-1}_{2k-1\,j} = C^{2l}_{2k\,j}\qquad 
C^{2l-1}_{2k\,j} = -C^{2l}_{2k-1\,j} \en$$
for $1 \le k,l \le p$ and $2p+1 \le j \le m$ and
$$C^k_{2i-1\,2j-1} = C^k_{2i\,2j} \qquad 
C^k_{2i-1\,2j} = -C^k_{2i\,2j-1} \en$$
for $1 \le i,j \le p$ and $k=0$ or $2p+1 \le k \le m$.
If $n_h = n_i +n_j$, the solutions of equations (17a,b) are 
$$\eqalignno{
\quad C^{2h-1}_{2i-1\,2j-1} = -C^{2h-1}_{2i\,2j} = \quad C^{2h}_{2i-1\,2j} 
&= \quad C^{2h}_{2i\,2j-1}\enpt \cr
\quad C^{2h-1}_{2i-1\,2j} \quad = C^{2h-1}_{2i\,2j-1} = -C^{2h}_{2i-1\,2j-1} 
&= \quad C^{2h}_{2i\,2j}\endpt
}$$
whereas if $n_h =n_i - n_j$ the solutions are
$$\eqalignno{
-C^{2h-1}_{2i-1\,2j-1} = -C^{2h-1}_{2i\,2j} = -C^{2h}_{2i-1\,2j} 
&= \quad C^{2h}_{2i\,2j-1}\enpt \cr
-C^{2h-1}_{2i-1\,2j} \quad =  C^{2h-1}_{2i\,2j-1} = \quad C^{2h}_{2i-1\,2j-1} 
&= \quad C^{2h}_{2i\,2j}\endpt
}$$
and for $n_h =n_i - n_j$ the solutions are obtained from (22) by
switching $i$ and $j$.

The structure constants must also satisfy the Jacobi identities 
$$[X_i\ [X_j\ X_k]] + [X_j\ [X_k\ X_i]] +[X_k\ [X_i\ X_j]] = 0 \en $$
where $i,j,k = 1 \ldots m$. However in the case general these do not lead
to particularly simple relations and so will not be considered at this
stage.  However this identity is used in the analysis of the four
dimensional Lie groups below.

\section{Cyclically symmetric manifolds admitting a $G_2$ or $G_3$}
Before considering the four dimensional case the results of the
previous section are applied to the two and three-dimensional cases.
For the two-dimensional case ($m=1$) equation (5) reduces to 
$[X_0\ X_1] = 0$ and so the group is Abelian which is the result of
Bi\v c\'ak \& Schmidt (1984). 

For the three-dimensional case ($m=2$)
equation (5) allows two possibilities: either $X_1$ and $X_2$ commute
with $X_0$ or we have $[X_0\ X_1] = n X_2$ and $[X_0\ X_2] = -nX_1$
for some positive integer $n$.  In the former case the Jacobi
identities are identically satisfied  and by a basis transformation of
the form (6c) we may reduce the commutation relations
to one of three following algebraically distinct forms
$$\eqalignno{
\fl  \qquad {\rm Bianchi\ type\ I} \qquad
& [X_0\ X_1] = 0\qquad [X_0\ X_2] = 0 \qquad [X_1\ X_2] = 0 \cr
\fl \qquad {\rm Bianchi\ type\ II} \qquad
& [X_0\ X_1] = 0\qquad [X_0\ X_2] = 0 \qquad [X_1\ X_2] = X_0 \cr
\fl \qquad {\rm Bianchi\ type\ III} \qquad
& [X_0\ X_1] = 0\qquad [X_0\ X_2] = 0 \qquad[X_1\ X_2] = X_2
}$$
In the latter case equation (9) implies $C^1_{12} = C^2_{12} = 0$ and
so $[X_1\ X_2] = a X_0$ where $a$ is a constant.  
By using the freedom to rescale $X_1$ and $X_2$ as in equations (6a,b) we
may set $a=0,\pm 1$ and so arrive at one of the three algebraically
distinct forms
$$\eqalignno{ \fl
\qquad {\rm Bianchi\ type\ VII}_0 \qquad 
& [X_0\ X_1] = nX_2 \qquad [X_0\ X_2] = -nX_1 \qquad [X_1\ X_2] = 0 \cr
\fl \qquad {\rm Bianchi\ type\ VIII} \qquad 
& [X_0\ X_1] = nX_2 \qquad [X_0\ X_2] = -nX_1 \qquad [X_1\ X_2] = -X_0 \cr
\fl \qquad {\rm Bianchi\ type\ IX} \qquad
& [X_0\ X_1] = nX_2 \qquad [X_0\ X_2] = -nX_1 \qquad [X_1\ X_2] = X_0
}$$
We may set $n=1$ in the second set of
commutators by rescaling $X_0$ and so obtain 
the canonical forms of the commutation relations for Bianchi types
VII$_0$, VIII and IX\null.  However, if this is done, we cannot
maintain both the $2\pi$ periodicity of the coordinate $\phi$ and the
relation $X_0 = \partial_\phi$. 

\section{Cyclically symmetric manifolds admitting a $G_4$}
For the four-dimensional case ($m=3$)
equation (5) allows two possibilities: either $X_1$, $X_2$ and $X_3$
all commute with $X_0$ or we have 
$$[X_0\ X_1] = n X_2 \qquad [X_0\ X_2] = -nX_1 \qquad [X_0\ X_3] = 0
\en $$
for some positive integer $n$.  

In the former case the Jacobi identities involving $X_0$ are
identically satisfied. Thus the cyclic symmetry does not
imply further restrictions on the structure constants in this case.
The basis vectors $X_1$, $X_2$ and $X_3$ are only defined up
to the transformations of the form 
$$\tilde X_k = A^l_k X_l  + B_k X_0 \en $$
where $A^l_k$ and $B_k$ are constants for $k,l = 1 \ldots 3$
 The Jacobi identity (23) remains to be satisfied; this
leads to the following restrictions on the structure constants 
$$C^\alpha_{1l}C^l_{23} + C^\alpha_{2l}C^l_{31} + C^\alpha_{3l}
C^l_{12} = 0 \en$$ 
where $l= 1 \dots 3$ and $\alpha = 0 \ldots 3$.  The following
classification method may then be 
used:  choose $A^l_k$ in equation (25) so that $C^i_{jk}$ take one of
the nine canonical (Bianchi) forms for a three-dimensional Lie
algebra; then use equation (26) with $\alpha = 0$ and the freedom in
the choice of the basis vectors in equation (25) (choice of $B_k$ and
any remaining freedom in the choice of $A^l_k$) to simplify $C^0_{ij}$
as far as possible.  Although the analysis 
is straightforward, it will not be presented here as more than a dozen
separate cases need to be considered.

In the latter case where equation (24) is satisfied, the Jacobi
identities involving $X_0$ imply that 
$$[X_1\ X_2] = a X_3 +b X_0 
\qquad [X_2\ X_3] = c X_1 +d X_2 
\qquad [X_3\ X_1] = -d X_1 + c X_2$$
where $a$, $b$, $c$ and $d$ are constants.  However, without loss of
generality, we may set 
$c=0$ by means of the basis change 
$\tilde X_3 = X_3 - c/n X_0$.
Futhermore the remaining Jacobi identity (23) implies $d (aX_3 + b
X_0) = 0$ so that either $d=0$ or $a = b = 0$. 

If $d \ne 0$, we may set $d = 1$ by rescaling $X_3$ and the commutation
relations take the form 
$$\eqalign{  
& [X_0\ X_1] = nX_2 \qquad [X_0\ X_2] = -nX_1 \qquad [X_0\ X_3] = 0 \cr
& [X_1\ X_2] = 0 \qquad [X_2\ X_3] = X_2 \qquad [X_3\ X_1] = -X_1
}$$
which, apart from the numbering and signs of the basis vectors, is the
canonical form for 
Kruchkovich-Petrov class V (Kruchkovich, 1954 \& 1957, Petrov 1963).
There is no three-dimensional sub-algebra containing $X_0$, but clearly
there is a sub-algebra spanned by $X_1$, $X_2$ and $X_3$. 

If $d = 0$, then the commuation relations take the form
$$\eqalign{  
& [X_0\ X_1] = nX_2 \qquad [X_0\ X_2] = -nX_1 \qquad [X_0\ X_3] = 0 \cr
& [X_1\ X_2] = a X_3 + b X_0 \qquad [X_2\ X_3] = 0 \qquad [X_3\ X_1] = 0
}$$
Note that $X_3$ generates the centre of the Lie algebra, that is it is
the only vector that commutes with all the vectors of the algebra.
There are essentially only six algebraically distinct cases: 

\vskip 9pt
\noindent (i)  If $a = b = 0$,  $X_1$, $X_2$ and $X_3$ generate an
abelian sub-algebra and so the Lie algebra is of 
Kruchkovich-Petrov type VI; in fact it is of type VI$_4$ with $k=l=0$
(Petrov 1963).  This algebra is a central extension by $X_3$ of the
Bianchi type VII$_0$ algebra in section 3. Here and in classes
(ii)-(iv) below, the vectors $X_0$, $X_1$, $X_2$ and $X_3$ form a 
canonical basis of the Lie algebra. 

\vskip 9pt
\noindent  (ii) If $a=0$ and $b=+1$, the algebra is of
Kruchkovich-Petrov type VIII.  This algebra is a central extension by
$X_3$ of the Bianchi type IX algebra in section 3.

\vskip 9pt
\noindent (iii) If $a=0$ and $b=+1$, the algebra is of
Kruchkovich-Petrov type VII.  This algebra is a central extension by
$X_3$ of the Bianchi type VIII algebra in section 3. 

\vskip 9pt
\noindent (iv)  If $a=1$ and $b=0$, the algebra is of
Kruchkovich-Petrov type III$_{q=0}$.   There is no three-dimensional
sub-algebra containing $X_0$, but clearly there is a sub-algebra
spanned by $X_1$, $X_2$ and $X_3$.  

\vskip 9pt
\noindent (v)  If $a=b=1$, the algebra is again of
Kruchkovich-Petrov type VIII, but unlike class (ii), $X_0$ is not a
canonical basis vector.  Instead in this case a canonical basis of the
algebra is $X_0 + X_3$,   $X_1$, $X_2$ and $X_3$. 

\vskip 9pt
\noindent (vi)  If $a=1$ and $b=-1$, the algebra is of
Kruchkovich-Petrov type VII, but unlike class (iii), $X_0$ is not a
canonical basis vector. Instead in this case a canonical basis of the
algebra is $X_0 - X_3$,   $X_1$, $X_2$ and $X_3$. 

In classes (v) and (vi),  there is no three-dimensional
sub-algebra containing $X_0$, but clearly there is a sub-algebra
spanned by $X_1$, $X_2$ and $X_0 \pm X_3$ which is of Bianchi type IX
or VIII in cases (v) and (vi) respectively.

\section{Summary}
Lie transformation groups which act on an $n$-dimensional
manifold $\cal M$ and which contain a one-dimensional subgroup acting
cyclically on $\cal M$ have been considered.  It is found that the
existence of a cyclic action imposes quite severe restrictions on the
structure constants of the group. 

Such two-dimensional groups must be Abelian and for the
three-dimensional case the 
Bianchi type of the group is restricted: types IV, V and VI cannot
occur and  only the subclass VII$_{q=0}$ of type VII is allowed. The
results for the three-dimensional case were obtained by the author
(Barnes, 2000).   

At first sight the restrictions in the three-dimensional case  
may not seem too severe: only three of the nine Bianchi types are
excluded whilst a fourth type is restricted.  However, it should be
pointed out that 
Bianchi types VI and VII both involve an arbitrary parameter and so
each contain an infinite number of algebraically distinct types. Those
in type VI are excluded completely and for type VII only a
single case survives. Moreover for most of the types which {\it
are\/} permitted 
the structure of the Lie algebra is `aligned nicely' relative to $X_0$.
For example, for Bianchi type II, $X_0$ is also the generator of the
first derived subalgebra and for Bianchi types II and III, $X_0$ lies in
the centre of the algebra.   For Bianchi types VII$_0$, VIII and IX, $X_0$
is a member  of a canonical basis of the Lie algebra.

The structure of four-dimensional groups is again severely restricted
by the existence of a one-dimensional subgroup acting cyclically on
$\cal M$.  In the case when the generator of the cyclic action is not
in the centre of the algebra, only five  of the classes of Kruchkovich
(1954, 1957) and Petrov (1969) can occur: namely III, IV$_4$, V, VII
and VIII, with only zero-parameter sub-cases of types III and IV$_4$
occurring.  Moreover in all cases $X_0$ is `aligned nicely' relative to a
canonical basis of the Lie algebra.

\references 
\refjl{Barnes A 2000}{\CQG}{17} {2605--9}
\refjl{Bi\v c\'ak J and Schmidt B G 1984}{\JMP}{25} {600--6}
\refjl{Carot J, Senovilla J M M and Vera R 1999}{\CQG}{16} {3025--34}
\refjl{Carter B 1970}{Commun. math. Phys.}{17}{233--8}
\refjl{Kruchkovich G I 1954} {Usp. Mat. Nauk.}{9}{part 1 (59), 3}
\refjl{Kruchkovich G I 1957} {Mat. sbor.}{41}{(83), 195}
\refjl{Mars M and Senovilla  J M M 1993}{\CQG}{10}{1633--47}
\refbk{Petrov A Z 1969}{Einstein Spaces}{Pergamon Press, Oxford, p.~63}

\bye